\documentclass[prb,letterpaper,twocolumn,showpacs]{revtex4} 
\usepackage{times,xspace} 
\usepackage{amsbsy,amssymb,amsmath,bm} 
\usepackage{graphicx,color,epsfig,rotate} 
\usepackage{fancyhdr} 
 
\def\bbbc{{\mathchoice {\setbox0=\hbox{$\displaystyle\rm C$}\hbox{\hbox 
to0pt{\kern0.4\wd0\vrule height0.9\ht0\hss}\box0}} 
{\setbox0=\hbox{$\textstyle\rm C$}\hbox{\hbox 
to0pt{\kern0.4\wd0\vrule height0.9\ht0\hss}\box0}} 
{\setbox0=\hbox{$\scriptstyle\rm C$}\hbox{\hbox 
to0pt{\kern0.4\wd0\vrule height0.9\ht0\hss}\box0}} 
{\setbox0=\hbox{$\scriptscriptstyle\rm C$}\hbox{\hbox 
to0pt{\kern0.4\wd0\vrule height0.9\ht0\hss}\box0}}}}

\newcommand{\ignore}[1]{} 
\newcommand{\pComment}[1]{} 
\newcommand{\gComment}[1]{} 
\newcommand{\jComment}[1]{} 
\newcommand{\rComment}[1]{} 
\newcommand{\lComment}[1]{} 
 
\renewcommand{\pComment}[1]{\textcolor{blue}{Pinaki: #1}} 
\renewcommand{\gComment}[1]{\textcolor{red}{Gerardo: #1}} 
\renewcommand{\jComment}[1]{\textcolor{green}{Jim: #1}} 
\renewcommand{\rComment}[1]{\textcolor{magenta}{Ray: #1}} 
\renewcommand{\lComment}[1]{\textcolor{purple}{Rolando: #1}} 
 
\begin{document}

\title{Pairing Correlations in the two-layer attractive Hubbard Model} 

\author{Aleksander Zujev$^{1}$,  Richard T.~Scalettar$^{2}$, 
George~G.~Batrouni$^{3,4,5}$ and Pinaki~Sengupta$^{1}$}

\affiliation{$^{1}$School of Physical and Mathematical Sciences, Nanyang Technological
University, 21 Nanyang Link, Singapore 637371 \\
$^{2}$Department of Physics, University of California, Davis,
California 95616-8677 \\
$^{3}$Institut Non-Lin\'eaire de Nice, UMR 6618 CNRS,
Universit\'e de Nice--Sophia Antipolis, 1361 route des Lucioles,
06560 Valbonne, France\\
$^{4}$Institut Universitaire de France\\
$^{5}$Centre for Quantum Technologies, National University of
Singapore; 2 Science Drive 3 Singapore 117542}

\date{\today} 

\begin{abstract}
  Studies of systems with two fermionic bands with {\it repulsive}
  interaction strength $U$ have a long history, with the Periodic
  Anderson Model (PAM) being one of the most frequently considered
  Hamiltonians.  In this paper, we use Quantum Monte Carlo to study
  analogous issues for {\it attractive} interactions.  As in the
  Periodic Anderson Model, we focus on a case where one band is
  uncorrelated ($U=0$), and focus on the effect of hybridization $V$
  between the bands on the pairing correlations.  A key difference
  with the PAM is that there is no sign problem, so that we are able
  to explore the physics of {\it doped} multi-band attractive systems
  at low temperatures whereas ground state properties of repulsive
  models can be determined only at half-filling.  For small $V$,
  pairing in the $U<0$ layer induces pairing in the $U=0$ layer.  At
  larger $V$ the ground state of the coupled system loses its
  superconducting character.  The Quantum Monte Carlo data are
  complemented by results obtained with the Bogoliubov-de Gennes
  approximation.
\end{abstract}


\maketitle
\thispagestyle{fancy}

\section{Introduction}

Cuprate superconductors are characterized by CuO$_2$ planes which are
rather isolated from each other by intervening rare earth atoms.  The
larger (``$c$-axis") separation of Cu atoms perpendicular to the
planes compared to the in-plane lattice constants ($a,b$) has focused
theoretical attention on the magnetic and superconducting properties
of two-dimensional models, most notably the square lattice Heisenberg
and Hubbard Hamiltonians.  Indeed, one of the most fundamental
theoretical questions which arose in the initial investigations of
possible models of high temperature superconductivity concerned the
nature of the magnetism in the ground states of these Hamiltonians:
Were they spin-liquids or did they exhibit long range order?  It was
established numerically\cite{reger88,hirsch89,white89} that both the
Heisenberg Hamiltonian and the half-filled Hubbard Hamiltonian have
long range antiferromagnetic order at $T=0$.  A still-unresolved
question is the nature of pairing in the doped Hubbard Hamiltonian.

A natural extension of these single layer questions concerns the
behavior of spin and pairing correlations in coupled planes.
Interlayer connections are required to elevate the magnetic ordering,
which is possible only at $T=0$ in two dimensions, to the finite
temperature Ne\'el transitions observed experimentally.  Similarly,
the superconducting transition temperature tends to increase with the
number of adjacent CuO$_2$ layers between the charge resevoirs,
getting steadily larger from single layered La$_{2-x}$Sr$_x$CuO$_4$,
to bilayer YBaCu$_3$O$_{6+y}$ to the HgBaCaCuO sequence where $T_c$
peaks at about $130^\circ$ K for materials with $n=3$ layers
\cite{shimizu12}.  Antiferromagnetism and thus superconductivity
exhibit the expected behavior that adding additional (non-frustrating)
couplings, and increasing the dimensionality, enhances the tendency to
order.

The route from 2D to 3D is, however, not completely straightforward.
Quantum Monte Carlo (QMC) studies of coupled Heisenberg
layers\cite{sandvik94}, have shown that when the interplane coupling
grows sufficiently large, there is a Quantum Phase Transition (QPT)
from a ground state with long range magnetic order to a 
 spin-gapped ground state,
in which singlets form between pairs of spins in the two
planes\cite{foot1}.  A similar effect is seen in QMC studies of the
two-layer half-filled Hubbard Hamiltonian \cite{scalettar94}.

The basic phenomenon of singlet formation usurping magnetic order in
itinerant fermion Hamiltonians does not crucially depend on having the
same interaction strength $U$, or hopping $t$, for the two species.
Indeed, a Hamiltonian with $U=0$ in one (conduction) band and hopping
$t=0$ in another (localized) band, is the commonly encountered
Periodic Anderson Model (PAM).  At half-filling, for weak
hybridization between the conduction and localized bands,
antiferromagnetic order is present.  As the hybridization grows
bigger, however, ``Kondo" singlets form and destroy antiferromagnetic
(AF) order.  Thus the PAM exhibits a similar QPT as for two identical
Hubbard layers ($t_1=t_2$ and $U_1=U_2$).  This close similarity of
the PAM and Hubbard bilayers emphasizes that the physics of fermions
in multiple layers can equivalently be interpreted in in terms of
multiple orbitals.

The purpose of this paper is to study the possibility of analogous
phenomena in the case when there is an {\it attractive} interaction.
More specifically, we examine the nature of
pairing correlations for coupled planes (orbitals) in situations when an
attraction is present only in one plane (orbital).  We use the $-U$
Hubbard Hamiltonian as an appropriate simple model, and employ a
combination of Quantum Monte Carlo (QMC) and Bogoliubov-de Gennes (BdG)
Mean Field Theory.  We will focus on a case analogous to that of the
PAM, namely when $U=0$ in one of the layers.  We are interested in both
how the interlayer hopping affects the $s$-wave pairing correlations in
a layer with $U<0$, and also whether pairing can be induced by a
``proximity effect" in the $U=0$ layer.  Such behavior would be directly
analogous to the ``RKKY" spin polarization cloud which is induced in the
conduction band by the presence of the local moments in the PAM.  A
particularly interesting issue is whether the QPT which occurs at
commensurate filling also occurs in the doped case.

The remainder of this paper is organized as follows: In Sec.~II we write
down the precise model and give an overview of the QMC and BdG
computational methodologies.  Secs.~III and IV present the BdG and QMC
results respectively.  Sec.~V recaps our conclusions.

\section{Model and Computational Methods}

Our starting point is the two layer (orbital) attractive Hubbard Hamiltonian,
\begin{eqnarray}
\label{eq:attr-hub-mod}
H=&-&t\sum_{<ij> l \sigma}
( c^{\dagger}_{i l \sigma}c^{\phantom{\dagger}}_{j l \sigma} +
  c^{\dagger}_{j l \sigma}c^{\phantom{\dagger}}_{i l \sigma})
\nonumber \\
&-&V \sum_{i \sigma}
( c^{\dagger}_{i 1 \sigma}c^{\phantom{\dagger}}_{i 2 \sigma} +
  c^{\dagger}_{i 2 \sigma}c^{\phantom{\dagger}}_{i 1 \sigma})
- \sum_{i l \sigma} \mu^{\phantom{\dagger}}_l c^{\dagger}_{i l \sigma}
c^{\phantom{\dagger}}_{i l \sigma}
\nonumber \\
&-&\sum_{i}\big|U^{\phantom{\dagger}}_{1}\big|
(n^{\phantom{\dagger}}_{i 1 \uparrow}-\frac12) 
(n^{\phantom{\dagger}}_{i 1 \downarrow}-\frac12)\,\,\,.
\label{eq:ham}
\end{eqnarray}
Here an intralayer kinetic energy term $c_{i l \sigma}^\dagger
c^{\phantom{\dagger}}_{j l \sigma}$ describes the creation of a
fermion with spin $\sigma$ on site $i$ of layer $l$ and its
destruction on site $j$ of the same layer.  A second interlayer
kinetic energy term $c^{\dagger}_{i l \sigma}c^{\phantom{\dagger}}_{i
  l^\prime \sigma}$ hybridizes the two layers $l,l^\prime$.  Fermions
of different spin feel an attractive interaction $-|U_l|$ on layer
$l$.  The chemical potentials $\mu_l$ control the filling.  Our
lattice geometry consists of two coupled square lattice of linear size
$L$.  We choose $U_1<0$ in the `superconducting' plane and $U_2=0$ in
the metallic one.

At half-filling, a particle-hole transformation formalizes the
similarity of the repulsive and attractive models.  The vanishing of
antiferromagnetic order with increasing $V$ for $U>0$ bilayers
immediately implies that ground state pairing (and charge density wave
order, with which it is degenerate) must be destroyed as $V$ grows for
$U<0$.  However, whereas AF order occurs only at half-filling (and
only at $T=0$) in the 2D repulsive Hubbard Hamiltonian,
superconductivity appears away from half-filling, and at a finite
(Kosterlitz-Thouless) transition temperature
\cite{scalettar89,moreo91}, in the attractive Hubbard Hamiltonian.
This opens up fundamentally new issues in the attractive case.
Whether the fermions on sites in adjacent layers will still lock into
a local object and destroy long range order when the filling is
incommensurate is an interesting question.  The absence of a sign
problem in the case of the attractive Hubbard model enables the study
of low temperatures even when the filling is incommensurate, allowing
access to ground state properties.

We perform ``determinant" QMC \cite{blankenbecler82} simulations of
the Hamiltonian, Eq.~\ref{eq:ham} by writing down a path integral for
the partition function $Z$.  The exponential of the interaction term
$U_l$ is decoupled with a Hubbard-Stratonovich field which allows the
trace of the remaining exponentials of quadratic forms of the fermion
operators to be performed analytically.  The result is a sum over
configurations of the Hubbard-Stratonovich field with a weight which
takes the form of the product of the determinants of two matrices of
dimension the spatial lattice size, one for each spin species.  In the
case of attractive interaction the two matrices, and hence their
determinants, are identical, so that the weight is a perfect square
and there is no sign problem.

We present results here for the real space pair correlation functions
in the two orbitals,
\begin{eqnarray}
P_l(j) &=& \langle \Delta^{\phantom{\dagger}}_{i+j,l}
\Delta^\dagger_{i,l} \rangle 
\nonumber \\
\Delta^\dagger_i &=& c^{\dagger}_{i l \uparrow} c^{\dagger}_{i l
  \downarrow}
\label{eq:paircorrspace}
\end{eqnarray}
and also their associated structure factors,
\begin{eqnarray}
{\cal P}^l_s = \frac{1}{N} \sum_{j} P_l(j)
\label{eq:pairstructure}
\end{eqnarray}
as a functions of interband hybridization $V$, on-site attraction
$U_l$, density $\rho$, and temperature $T$.  The long distance
behavior of $P^{\phantom{\dagger}}_l(j)$ yields the square of the
superconducting order parameter, as does the lattice size dependence
of the structure factor.

Eq.~\ref{eq:ham} can also be studied in Mean Field Theory via the
solution of the BdG equations.  In this approach, the four fermion
(interaction) term is decoupled in the Hamiltonian itself,
\begin{eqnarray}
  {\cal{H}}_{\rm eff}=&-&t\sum_{<ij>l\sigma}
  (c^{\dagger}_{il\sigma}c^{\phantom{\dagger}}_{jl\sigma} +
  c^{\dagger}_{jl\sigma}c^{\phantom{\dagger}}_{il\sigma}) 
  \nonumber \\
  &-&V \sum_{i \sigma}
  ( c^{\dagger}_{i 1 \sigma}c^{\phantom{\dagger}}_{i 2 \sigma} +
    c^{\dagger}_{i 2 \sigma}c^{\phantom{\dagger}}_{i 1 \sigma})
  - \sum_{i l \sigma}{\tilde\mu^{\phantom{\dagger}}_{i l}}
  c^{\dagger}_{i l \sigma}c^{\phantom{\dagger}}_{i l \sigma}
  \nonumber \\
  &-& \sum_{i}
  \big|U^{\phantom{\dagger}}_{1}\big|\big[
   \Delta^{\phantom{\dagger}}_{i1}
c^{\dagger}_{i 1 \uparrow}c^{\dagger}_{i 1 \downarrow}
 +\Delta^{*\phantom{\dagger}}_{i1}
  c^{\phantom{\dagger}}_{i 1 \downarrow}
  c^{\phantom{\dagger}}_{i 1 \uparrow}\big]\,\,\,.
\label{eq:HBdG}
\end{eqnarray}
Here the gap
$\Delta^{\phantom{\dagger}}_{il}=\big<c^{\phantom{\dagger}}_{il\uparrow}
c^{\phantom{\dagger}}_{il\downarrow}\big>$ and density
$\big<n^{\phantom{\dagger}}_{il\sigma}\big>=\big<c^{\dagger}_{il\sigma}
c_{il\sigma}\big>$ are determined self-consistently by diagonalizing
the quadratic BdG Hamiltonian and putting together the eigenvalues and
eigenvectors appropriately to compute refined values which are
inserted back in the Hamiltonian in an iterative process.  A related
BdG treatment of superconductivity in the presence of randomness
(spatially varying $\mu_i$) is given in
[\onlinecite{ghosal98,ghosal01}].
${\tilde\mu_{il}}=\mu+\big|U_{l}\big| \langle n_{il} \rangle/2$
includes a site-dependent Hartree shift with $\langle n_{il} \rangle
=\sum_{\sigma} \langle n_{il\sigma} \rangle$.

In the general case, when inhomogeneous terms are present in the
Hamiltonian, or are expected to develop spontaneously, (like the
charge and spin stripes of the doped, repulsive Hubbard model), the
order parameter and densities are allowed to depend on site index and
the diagonalization must be done numerically.  If translation
invariance is present, Eq.~\ref{eq:HBdG} can be diagonalized
analytically by going to momentum space.  The resulting momentum sums
are typically still done numerically, but larger lattices can be
studied.  We have implemented both approaches in the work presented
here.

Although inhomogeneous Hartree-Fock theory allows for spatial variation
of the densities and superconducting order parameters, 
the expectation values in Eq.~\ref{eq:HBdG} 
are independent of imaginary time, unlike the fluctuating
Hubbard-Stratonovich field in the QMC approach.  This leads to a less
accurate treatment of interparticle correlations, but
a numerically much more simple problem.  In particular, larger spatial
lattices can be studied, and the BdG solution can exhibit broken
symmetry so that $\Delta$ itself can be nonzero, rather
than having to be extracted from the asymptotics of
the correlation function Eq.~2.  Here the combination of QMC and
BdG will serve to provide complementary information.

In the case of the BdG calculations our focus will be on the value of
the order parameter $\Delta$ as a function of the
same parameters as varied in the QMC.  As mentioned above,
the asymptotic value of $P_l(j)$ is a measure 
of $\Delta^2$.  The BdG approach, since it
neglects fluctuations, yields larger values
of $\Delta$ than those of the QMC, which is an exact method.

\section{Bogoliubov-de Gennes Results}
\begin{figure}
\includegraphics[height=4in,width=2.5in,angle=-90]{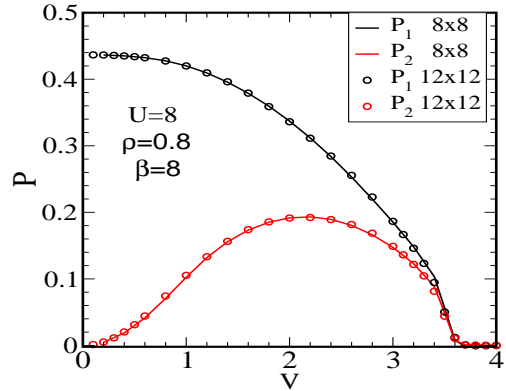}
\caption{ Bogoliubov-de Gennes result for $8\times 8$ and $12\times
  12$ bilayer lattices, $\beta=8$, $U_1=-8$, $\rho=0.8$.  The figure shows that
  there is an induced SC in the noninteracting layer, with negligible
  finite size effects.  }
\label{bfig:bdg} 
\end{figure}

The results of BdG calculations are summarized in
Fig.~\ref{bfig:bdg}. We consider two layers with $U_1/t=-8$ and
$U_2=0$, at a density of $\rho=0.8$ electrons per site and study the
evolution of the pairing correlation, $\Delta_l$, in the two layers as
a function inter-layer coupling, $V$. Pairing in the correlated layer
is suppressed by coupling to the non-correlated layer and eventually
destroyed completely at a critical $V_c$, analogous to the destruction
of antiferromagnetism by singlet formation in the repulsive case.  A
further interesting feature of Fig.~\ref{bfig:bdg} is that the $l=2$
layer, with $U_2=0$, develops induced pairing with the onset of
inter-layer coupling. The pairing amplitude, $P_2$, increases with increasing
$V$ up to a maximum before decreasing and eventually vanishing at
$V_c$. For $V> V_c$, the ground state consists of weakly interacting
inter-layer dimers. Crucially, the pairing order parameter in {\em
  both layers} remains non-zero over a finite range of inter-layer
coupling -- the induced pairing reported previously at half-filling
extends to finite dopings. In the next section, we present extensive
QMC results to confirm and complement the BdG results.

\section{Quantum Monte Carlo Results}

\subsection{Half Filling}

We begin our discussion of QMC results at half-filling,
$\rho_1=\rho_2=1$. As discussed previously, this corresponds to the
particle-hole symmetric point $\mu_1=\mu_2=0$ where there exists an
exact mapping to the repulsive Hubbard model and the results can be
benchmarked against previous studies of the Hubbard model
[\onlinecite{scalettar94,vekic95}]. Lattices of the form $2\times
L\times L$, with $4\leq L \leq 14$ were studied with periodic boundary
conditions over a wide range of inter-layer hybridization, and on-site
interaction for the interacting layer. An inverse temperature $\beta
t=L$ was found to be sufficient for the observables to have converged
to their ground state values.

\begin{figure}
\includegraphics[height=2.5in,width=3in,angle=0]{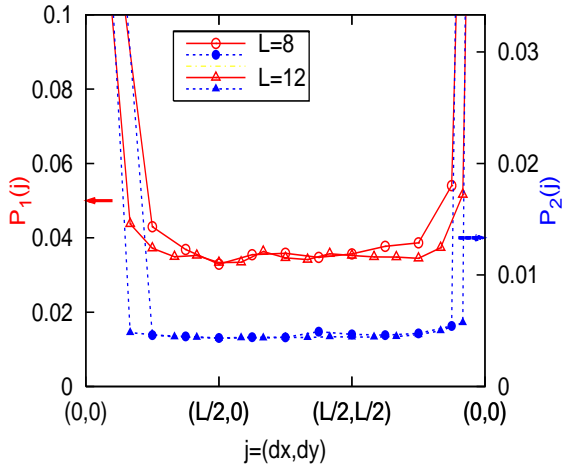}
\caption{The dependence of the ground state pair correlation
  functions, $P_{1,2}(j)$, on separation $j$ in a bilayer system
  comprised of one correlated layer coupled to an uncorrelated layer
  for two different lattice sizes $L$. The system is at half-filling,
  $\rho=1$, with interplane hybridization $V=0.8t$, and on-site
  interaction strength $U_1=-6t$ for the correlated layer.  The
  correlation functions converge to non-zero value at large
  separations, providing clear evidence for long-range order.  Despite
  the absence of interactions, there is proximity-effect induced
  long-range order in the uncorrelated layer, $l=2$, with an order
  parameter $\Delta_2$ approximately a factor of three lower than for
  the correlated layer.  As in Fig.~\ref{bfig:bdg}, finite size
  effects are modest.  }
\label{fig:P1db16D}
\end{figure}

The spatial dependence of the pair correlations $P_l(j)$ at
half-filling $(\rho=1)$, intersheet hybridization $V=0.75 t$, and
interactions $U_1=-6, U_2=0$ is shown in Fig.~\ref{fig:P1db16D}.  The
separation $j$ follows a trajectory along the $x$ axis to maximal $x$
separation $(\frac{L}{2},0)$ on a lattice with periodic boundary
conditions, and then to $(\frac{L}{2},\frac{L}{2})$ before returning
to separation $(0,0)$.  Results for lattices with $L=8, 12$ are shown
in the figure.  The interacting layer $l=1$ exhibits clear long range
order with $P_1(j)$ nearly independent of $j$ beyond a separation of
approximately $\frac{L}{4}$.  The noninteracting layer $l=2$ also
exhibits proximity-effect induced long range order, although the
correlation function is roughly an order of magnitude smaller.  This
corresponds to an order parameter $\Delta_l$ which is approximately a
factor of three smaller.

\begin{figure}
\includegraphics[height=3.5in,width=3in,angle=0]{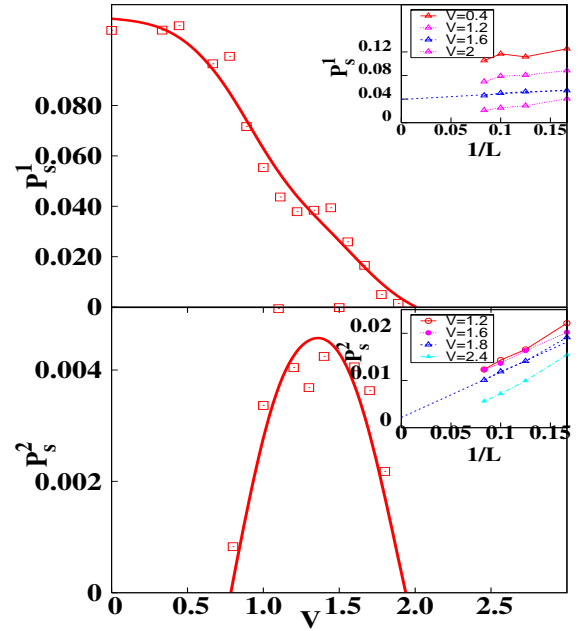}
\caption{The pair structure factor for a bilayer at half-filling for
  attractive on-site interaction in the correlated layer $U=-10t$. The
  main panels show the extrapolated structure factor as a function of
  the inter-layer coupling, $V$. The insets show the extrapolation of
  the data from finite-sized systems to the thermodynamic limit.  Top
  panel: The pair structure factor in the correlated layer, $P_s^{1}$.
  For inter-layer hybridization $V \le 1.6$ the extrapolated structure
  factor is non-zero indicating the existence of long-range order
  (LRO).  Bottom panel: The pairing structure factor $P_s^{2}$ in the
  non-interacting layer.  $P_s^{2}$ increases from zero for $V>0$,
  signaling the onset of induced pairing. With further increase of
  $V$, LRO vanishes simultaneously in both layers.
}
\label{fig:PFSS1} 
\end{figure}

Increasing the inter-layer hybridization reveals that the
pair structure factor for the interacting layer, $P^1_s$, decreases
monotonically and eventually the intra-layer pairing is destroyed at
some critical $V_c$. On the other hand, the pairing structure factor for
the non-interacting layer $P^2_s$ varies non-monotonically with $V$.  It
is vanishingly small at small $V$. As $V$ increases, long-range pairing
correlations increase, reach a maximum at an intermediate $V$ (which
depends on the strength of on-site interaction in the interacting layer)
and then decrease continuously to zero at  $V_c$. For $V>V_c$, the
ground state is dominated by {\it inter-}layer singlets.
Fig.~\ref{fig:PFSS1} shows the finite size dependence of the pairing
structure factor for the two layers at several values of $V$. For $V
\gtrsim1.6$, both $P^1_s$ and $P^2_s$ extrapolate to zero in the
thermodynamic limit indicating the absence of intra-layer pairing in
this regime. 

Here and further, the finite size scaling was done 
by linear fit in the plane $1/L \; - \; P^l_s$.
The scatter of the data in Fig.~\ref{fig:PFSS1} 
for different values of the interlayer hybridization $V$ 
is dominated by the error bars associated with the linear fit
to the structure factor on different lattice sizes used in the
extrapolation, as opposed to the statistical errors for
an individual run.  The data scatter serves as a stand-in
to provide a visual indication of the uncertainty in the
the calculation.

\subsection{Incommensurate Filling}

We now turn to the case of doped planes.  As mentioned in the
Introduction, this is interesting for two reasons.  First, there is a
possibility of a finite temperature KT transition because the CDW-SC
symmetry is broken.  Second, simulations of coupled, doped repulsive
layers at low $T$ are not possible.  Hence there is no DQMC
information available for the nature of magnetism in coupled layers
away from half-filling.  Our simulations address this issue, albeit
for attractive on-site interactions.

\begin{figure}
\includegraphics[height=3.5in,width=3in,angle=0]{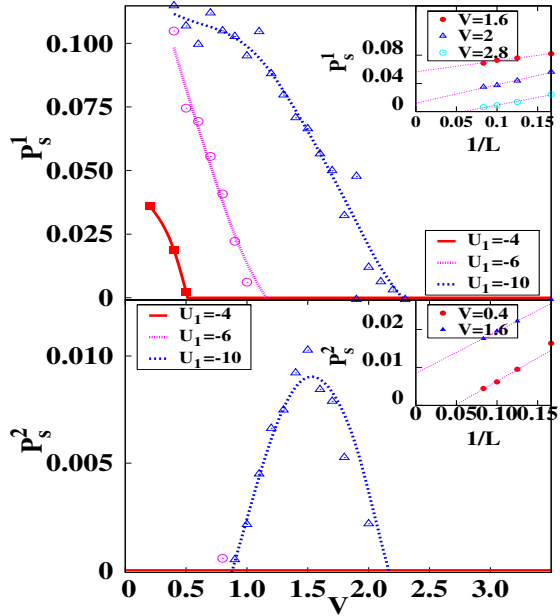}
\caption{The pair structure factor for a bilayer away from
  half-filling, $\rho=0.8$, for three representative values of the
  attractive on-site interaction in the correlated layer: $U_1=-4t,
  -6t, -10t$.  For $|U_1|\lesssim 6t$, $P_s^{2}=0$ at all values of
  $V$, indicating the absence of induced pairing. At stronger on-site
  interaction ($|U_1|>6t$) in the correlated layer, induced pairing
  appears in the uncorrelated layer over a finite range of $V$.  The
  critical $V_c$ for destruction of intra-layer pairing increases with
  $|U_1|$.  
  Insets: Finite size scaling for $U_1=-10t$.}
\label{fig:PFSS2} 
\end{figure}

The results in Fig.~\ref{fig:PFSS2} confirm unambiguously that induced
pairing in the uncorrelated layer extends to finite dopings away from
half-filling; indeed the pair correlation functions in the two layers
are qualitatively similar to those at half-filling. Pairing in the
correlated layer decreases monotonically to zero at a finite $V_c$ and
that in the non-interacting layer increases rapidly from zero as the
inter-layer hybridization is turned on, reaches a maximum value at
intermediate $V$ and then decreases to zero.  At intermediate to
strong values of the on-site interaction ($|U| \gtrsim 6t$), the
ground state of a system of coupled correlated and uncorrelated layers
away from half-filling consists of intra-layer pair formation in {\em
  both} layers over a finite non-zero range of inter-layer coupling.
Eventually, at sufficiently strong inter-layer hybridization, pairing
in both layers is destroyed and the ground state is dominated by
singlet formation between the layers. Induced pairing is found to
be absent for $|U| \lesssim 6t$. The simulation results are summarized
in Fig.~\ref{fig:PFSS2} where the extrapolated pairing structure
factor for the two layers are shown as a function of the inter-layer
hybridization, $V$, for some representative values of the on-site
interaction in the correlated layer at a fixed value of the density,
$\rho = 0.8$.  The maximum of the pairing structure factor in the
uncorrelated layer and the critical value of inter-layer coupling for
complete suppression of pairing, $V_c$, increase with $|U|$ . Our
results are consistent with the disappearance of pairing in both the
layers occurring simultaneously at $V_c$.  The error associated with
determining the extrapolated pairing structure factor in the
thermodynamic limit limits the accuracy with which $V_c$ can be
determined to about $\pm 0.2t$. In a marked departure from the
BdG results, the QMC data shows unambiguously that a finite non-zero
inter-layer coupling is required for the onset of induced pairing in the
uncorrelated layer. This is a consequence of the mean field nature 
of the BdG results.

Fig.~\ref{fig:PFSS2} shows that $P_1$ increases systematically
as $U$ changes from $U=-4$ to $U=-10$ and, indeed, shows no sign of saturation.
This is at variance with the behavior of the magnetic
response of the repulsive Hubbard Hamiltonian which first
increases with $U$ but then reaches a maximum at $U \sim 8t$
before falling.  This behavior is understood qualitatively
from the fact that the superexchange $J=4t^2/U$ 
which provides the large $U$ magnetic energy scale 
declines with $U$.  Thus, for example the Ne\'el temperature
of the 3D Hubbard model at half-filling is maximized at $U \sim 8t.$
The analog in the attractive Hubbard model is the fact that the
pairs become heavy, with an effective hopping 
$t_{\rm eff} \sim t^2/U$ associated with the fact that a pair
must be (temporarily) broken before it can hop.
This analogy suggests the pairing response might also
be maximal at an intermediate $U$.
However,
because of the sign problem, it is not known from DQMC whether such
a non-monotonic behavior exists in the repulsive model
away from half-filling.
Our results here for the attractive case, which show
no sign of saturation with increasing $U$, suggest that it might not.

\begin{figure}
\includegraphics[height=3.5in,width=3in,angle=0]{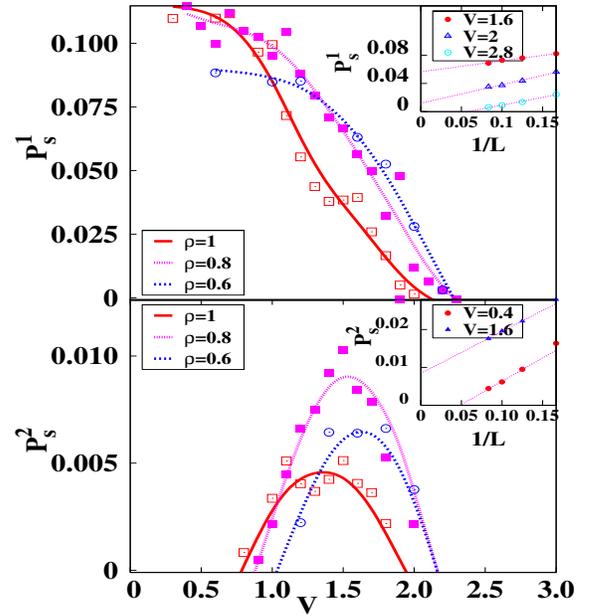}
\caption{The pair structure factors at various filling values.
  Induced pairing occurs at all densities, with the amplitude of
  induced pairing increasing with doping away from half-filling.  
  Insets: Finite size scaling for $\rho=0.8$. }
\label{fig:PFSS3} 
\end{figure}

Having established the existence of induced pair formation at finite
doping and the evolution of the associated structure factor with $V$,
we investigate the dependence of the pairing phenomenon on the density
of electrons in the layers. Our results (Fig.~\ref{fig:PFSS3}) show
that induced pairing occurs generically at all densities.  For all
density values studied, the pairing structure factor behaves in a
qualitatively similar manner, with the pairing strength optimally
enhanced at $\rho=0.8$, as evidenced by the initial increase in peak
height with decreasing density.  Pairing appears to show a modest
decline as the density is reduced further to $\rho=0.6$, although the
effect is rather small given the scatter in the data.  Such a
non-monotonic doping dependence would be consistent with the behavior
of the superconducting $T_c$ with doping for a single 2D sheet.  $T_c$
rises abruptly from zero at half-filling where charge density order
competes with pairing, but then shows a gradual decline below an
optimal doping \cite{scalettar89}.

\section{Conclusions}

In this paper we have used Quantum Monte Carlo and Bogoliubov-de Gennes
mean field theory to elucidate the properties of a multi-orbital {\it
  attractive} Hubbard Hamiltonian.  Our key conclusions are the
quantification of the induced pairing in a noninteracting orbital by
superconductivity in a correlated orbital, and, conversely, the
suppression of pairing in an orbital with an attractive interaction
through its coupling to a noninteracting orbital.

As we have emphasized in the introduction, analogous issues for
repulsive models have a long history.  In fact, over the last 4-5
years, the question of `orbitally selective Mott transitions', has
been extensively explored.  The fundamental objective has been an
examination of the density of states to determine whether or not one
subsystem can be metallic (zero energy gap at $E_{\rm F}$) and the
other insulating (nonzero gap), or whether the coupling forces the
transition to occur simultaneously in all orbitals (layers)
\cite{liebsch04,liebsch05,arita05,inaba06,costi07,koga04,ferrero05,demedici05,ruegg05,inaba05,knecht05,biermann05}.
Our work suggested that for attractive interactions the destruction of
superconductivity at large inter-orbital hybridization $V$ always
occurs simultaneously.  It is to be noted, however, that it appears
that at weak $V$ pairing can exist in the correlated orbital before it
is induced in the noninteracting one.  That is, we find that for the
attractive case at small interlayer coupling, SC exists only in the
interacting layer despite the fact that at larger $V$ the vanishing of
SC appears at a common $V_c$.


Our work is related to several other recent determinant QMC and
Lanczos studies of the attractive Hubbard model.  Assman {\it et.al}
\cite{assmann12} found that in an attractive Hubbard model with a
smoothly varying chemical potential (such as would be present in a
confined ultracold atomic cloud) pairing is significantly increased in
the half-filled portion of the lattice through its contact with doped
regions.  That is, the suppression of superconductivity caused by the
appearance of a degenerate charge density wave phase at $\rho=1$ is
eliminated.  Paiva {\it et.al.} explored a one-dimensional model of
borocarbides in which $U<0$ and $U=0$ sites alternate \cite{paiva03}.
Superconductivity is found to be possible only above a critical
density.

Finally we note that it was recently shown in repulsive models with
more than two layers that there can be further interesting and even
surprising evolution of magnetic properties with the hopping across an
interface between a Mott insulator and a metal \cite{euverte12}.  It
would be interesting to pursue related questions in the repulsive
model and, in particular whether the formation of a layer of local
pairs at the boundary can lead to a shielding of penetration effects.

We thank Simone Chiesa and Axel Euverte for very helpful dicsussions.
This work was supported in part by an ARO Award W911NF0710576 with
funds from the DARPA OLE Program, by the CNRS-UC Davis EPOCAL LIA
joint research grant; and by the CNRS-National University of Singapore
FSQL LIA joint research grant.

\end{document}